\documentclass[a4paper]{article}

\usepackage{INTERSPEECH2018}
\usepackage{xfrac}
\usepackage{url}
\usepackage{enumitem}
\usepackage{flushend}

\usepackage{xcolor}

\def\vec#1{\ensuremath{\bm{{#1}}}}

\def\tx#1{{\tt\footnotesize{#1}}}
\graphicspath{{./figures/}}
\DeclareGraphicsExtensions{.pdf,.jpg,.png}
\newcommand{\note}[1]{{#1}}

\title{A Weighted Superposition of Functional Contours Model for Modelling Contextual Prominence of Elementary Prosodic Contours}
\name{Branislav Gerazov$^{1,2}$, G\'erard Bailly$^2$ and Yi Xu$^3$}

\address{$^1$ FEEIT, University of Ss. Cyril and Methodius in Skopje, Macedonia \\
$^2$ Univ. Grenoble-Alpes, CNRS, Grenoble-INP, GIPSA-lab, 38000 Grenoble, France \\
$^3$ Division of Psychology and Language Sciences, University College London, UK}
\email{gerazov@feit.ukim.edu.mk, gerard.bailly@gipsa-lab.fr, yi.xu@ucl.ac.uk}

\begin{document}

\maketitle
\begin{abstract}
The way speech prosody encodes linguistic, paralinguistic and non-linguistic information via multiparametric representations of the speech signals is still an open issue. The Superposition of Functional Contours (SFC) model proposes to decompose prosody into elementary multiparametric functional contours through the iterative training of neural network contour generators using analysis-by-synthesis. Each generator is responsible for computing multiparametric contours that encode one given linguistic, paralinguistic and non-linguistic information on a variable scope of rhythmic units. The contributions of all generators' outputs are then overlapped and added to produce the prosody of the utterance. We propose an extension of the contour generators that allows them to model the prominence of the elementary contours based on contextual information. WSFC jointly learns the patterns of the elementary multiparametric functional contours and their weights dependent on the contours' contexts. The experimental results show that the proposed weighted SFC (WSFC) model can successfully capture contour prominence and thus improve SFC modelling performance. The WSFC is also shown to be effective at modelling the impact of attitudes on the prominence of functional contours cuing syntactic relations in French, and that of emphasis on the prominence of tone contours in Chinese.
\end{abstract}

\section{Introduction}

The way speech prosody encodes linguistic, paralinguistic and non-linguistic information via multiparametric representations of the speech signals is still an open issue. Most models of intonation postulate that this encoding is performed by local and salient spatio-temporal patterns such as tones, atoms or breaks inscribed into global gauges such as declinations or steps. Phonological structures are supposed to link socio-communicative functions with patterns and gauges. 

The Gestalt model proposed by Auberg\'e and Bailly~\cite{auberge1995generation} proposes that the encoding is direct, i.e. shapes make sense, and performed by spatio-temporal patterns that both cue each socio-communicative function and its scope, i.e. the linguistic units that are involved; e.g. the element carrying emphasis, the part of the utterance carrying doubt or the target syllable of a tone. The Superposition of Functional Contours (SFC) model developed by Bailly et al~\cite{bailly2005sfc, holm2002learning, bailly2002learning} bets that the parallel encoding of socio-communicative functions at multiple scopes is simply performed by overlapping-and-adding the function-specific spatio-temporal patterns.
The problem of decomposing prosody into these elementary patterns is ill-posed since the SFC does not impose any a priori constraints of the spatio-temporal patterns such as bandwidth or shape.

These function-specific patterns, in fact, emerge from statistical modelling. Given a dataset that contains multiple instances of these patterns, the SFC extracts the shapes and their average contributions thanks to an iterative analysis-by-synthesis training process that consists of training function-specific pattern generators. 
They are called {\em multiparametric contours} because the generated shapes feed a multiparametric score, i.e. one including melody, rhythm, head motion, etc.
The SFC has been successfully used to model different functions acting at various linguistic levels, including: attitudes~\cite{morlec2001generating}, grammatical dependencies~\cite{morlec1998evaluating}, cliticisation~\cite{bailly2002learning}, focus~\cite{brichet2004prosodie}, as well as tones in Mandarin~\cite{chen2004superposed}.

One shortcoming of the SFC model is that it is not sensitive to prominence: prosodic contours are simply superposed-and-added with no possibility of weighting their contributions. 
In this paper we supplement the SFC architecture with components responsible for weighting the contribution of the elementary contours in the decomposition. The weighted SFC (WSFC) consists in adjoining a {\em weight} module to each contour generator: while the contour generator still computes a multiparametric contour for each rhythmic unit of the scope, the weight module computes its contribution given the context of the scope in the utterance.

We assessed the plausibility of the proposed WSFC, and used it to explore two prosodic phenomena: \emph{i}) the impact of the attitude, and \emph{ii}) the impact of emphasis on the prominence of the other functional contours in the utterance. The two phenomena were explored in two different languages: French and Chinese. The results show that the integration of weighting in the contour generators is effective, relevant and robust. Also, by adding a degree of freedom to the model, it improves its modelling performance by providing more coherent contours. 
The whole implementation of the system has been licensed as free software and is available on GitHub\footnote{\footnotesize{\url{https://github.com/bgerazov/WSFC}}}.

\section{The SFC contour generators}

At the core of the SFC model are neural network based contour generators (CGs) that learn the elementary prosodic contours during the iterations of the analysis-by-synthesis loop 
\cite{morlec1997phd, morlec2001generating, holm2000generating, holm2002learning}. There is a single CG for each communicative function 
used in the dataset. 
Within an utterances, instances of these CGs are applied on the different scopes the functions encompass, i.e. the different number of rhythmic units (RUs), e.g. syllables, that the function spans. 

An example SFC decomposition of the intonation of a French utterance is shown in the left plot in Fig.~\ref{fig:wsfcdecomp}, where we can see a declaration contour overlapped with one left dependency between the verbal group and its subject, one right dependency between the verbal phrase and its direct object, and two clitic contours cueing articles. The decomposition was performed using the PySFC
prosody analysis system \cite{gerazov2018pysfc}. 

\section{The WSFC contour generators}

\begin{figure}[t]
 \centering
 \includegraphics[width=.5\linewidth]{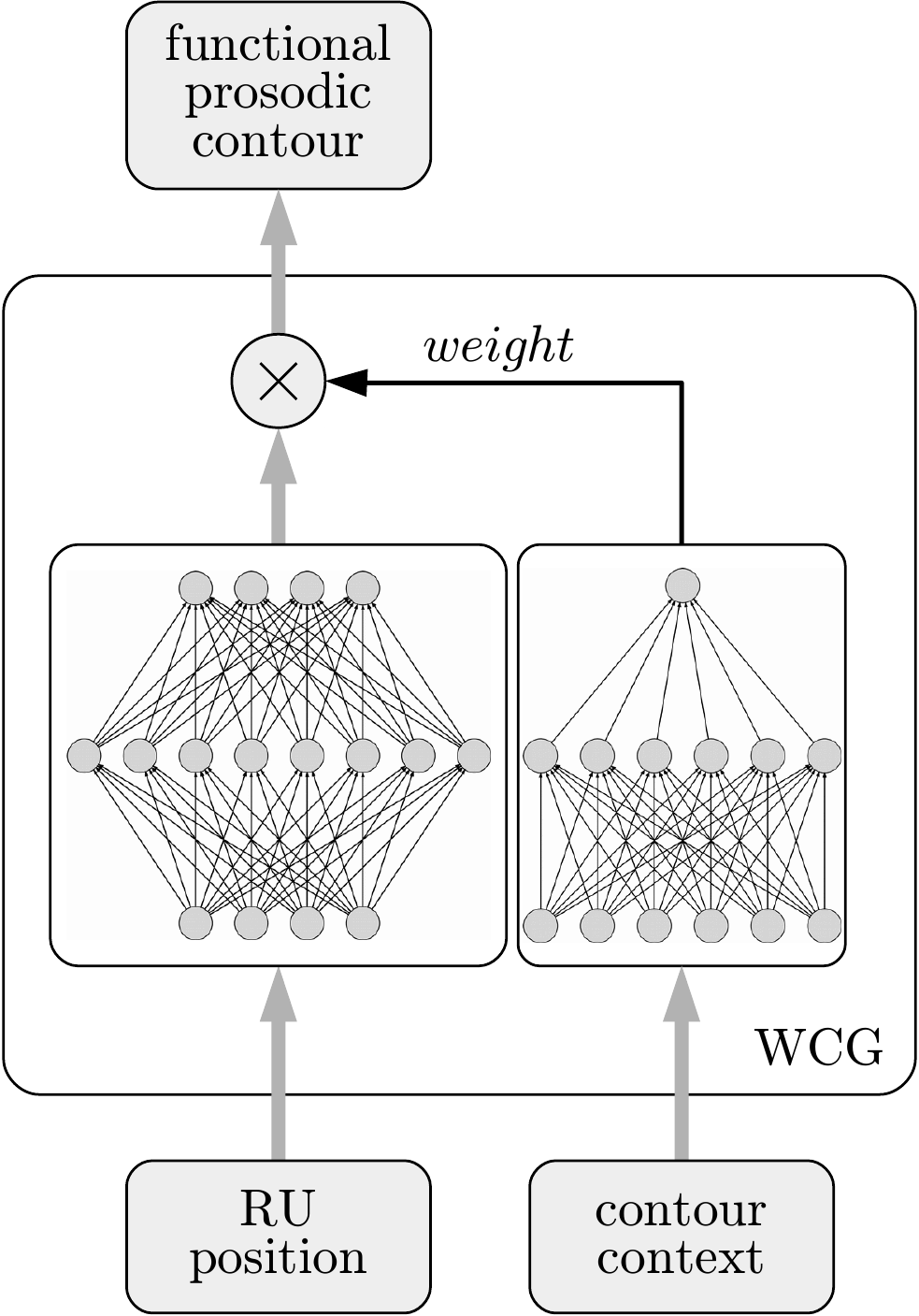}
 \vspace*{-0.2cm}
 \caption{Weighted contour generator introduced in the WSFC that features the SFC contour generator (left) gated by the weighting module (right).}
  \label{fig:wsfcc}
\end{figure}

\note{The WSFC is based on the weighted contour generator (WCG) shown in Fig.~\ref{fig:wsfcc}, which is an expanded form of the CGs introduced in the SFC. The WCG includes a module for computing the contour's weight that is designed to capture prominence based on linguistic context.} The global architecture is reminiscent to the {\em mixture of experts} (ME) model proposed by Jacobs et al~\cite{jacobs1991adaptive} in which predictions of experts, in our case contour generators, are weighted by gates, in our case weighting modules, and added to perform decisions or regressions.

\note{The contour generator module, comprising a single layer neural network, receives the RU's absolute and relative positions within the function's scope, and generates the prosodic contour for that particular syllable. The weight module, also comprising a shallow neural network, receives a vector describing the linguistic context that the functional contour appears in. Based on this context vector it outputs a global context-specific weight for all of the RUs within the function's scope. By imposing a single weight for the whole scope of the contour we force the weighting module to capture the overall prominence of the functional contours dependent on their context.} 
The input context vector can be arbitrarily defined and tailored to the task at hand. \note{In our experiments, we use three different binary encodings of the coinciding functional contours as context vectors to analyse the impact of attitude and emphasis on the prosodic contours; more details are given in Section \ref{sec:exp}.}

The two modules in the WSFC CG are in mutual competition, in the sense that the general amplitude of the contour generated by the contour generator is multiplied by the output of the weighting module. In that sense, the amplitudes of the contours can become arbitrarily large if they are compensated by the weights becoming small. To limit this effect the range output by the weight module is limited to 0 -- 2 \note{through multiplying with 2 the final neuron's sigmoid output.}
In addition, we apply regularisation to the weights' mean across the data towards $1.0$. This intuitively corresponds to an average function contour generated by the SFC model. Since training of the CGs is done in batches, the batch size is an important factor to take into account in this regularisation. 

\section{Objective and goals}
Our objective is to assess the efficiency of the WSFC for modelling the prominence of the elementary prosodic contours. We hypothesise that \emph{i}) the added degree of freedom will increase the modelling performance compared to the original SFC model. 
We will also use the WSFC to analyse the impact of linguistic context on the prominence of prosody contour realisation in a structured way. In this sense we will test an additional hypothesis: \emph{ii}) the WSFC is able to capture the impact that function contour context has on their prominence.

\section{Experiments}
\label{sec:exp}
We conducted three experiments to confront the WSFC with empirical data.

\subsection{Databases}
\skip -0.4cm
Three databases are used in the experiments:
\setlist{nolistsep}
\begin{itemize}[noitemsep]
\item \tx{Morlec} -- a database of 6 attitudes in French: declarative, question, exclamation, incredulous question, suspicious irony and obviousness~\cite{morlec2001generating}. In total, 1925 utterances are recorded from a single speaker with a total of 7956 syllables,
\item \tx{Liu} -- a database of declarations and five types of questions (y/n, wh~\ldots) in Chinese 
that include emphasis at three different positions.
The four carrier sentences are each built with words using one of the four Chinese tones.
The database comprises 76 sentences that are repeated 5 times by 6
speakers \cite{liu2005parallel}. In our work we use the subset from the first female speaker with a total of 380 utterances and 3820 syllables, and
 \item \tx{Chen} -- a database of read Chinese from a single speaker comprising 108 carrier utterances ranging from 6 to 38 syllables in length, with a total of 3470 syllables~\cite{chen2004superposed}.
\end{itemize}

\begin{figure*}[t]
 \centering
 \includegraphics[width=\linewidth]{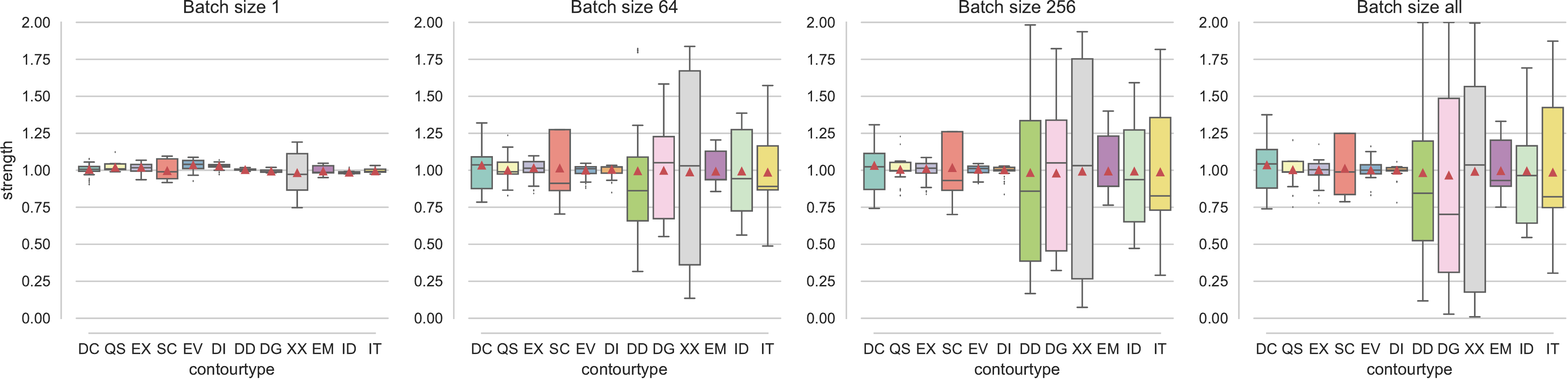}
 
 \vspace*{0.2cm}
 
 \includegraphics[width=\linewidth]{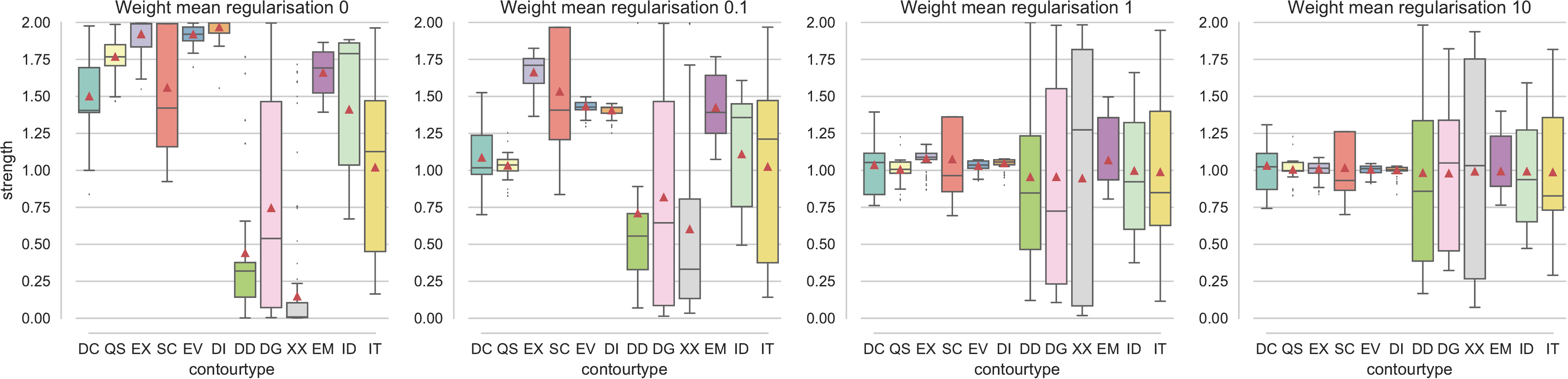}
 \vspace*{-0.5cm}
 \caption{Plots of the weights distribution in~\tx{Morlec} for varying batch sizes and a regularisation coefficient of 10 (top row) and for varying regularisation coefficients and a batch size of 256 (bottom row). The 12 functions in the database are: declaration (DC), question (QS), exclamation (EX), incredulous question (DI), suspicious irony (SC), obviousness (EV), dependency to the right and left (DD and DG), clitic (XX), emphasis (EM), independence (ID) and interdependence (IT).}
\label{fig:strengthsperbatch}
\end{figure*}

\subsection{Weight hyperparameters}
\skip -0.4cm
The two most important hyperparameters of the weight mechanism in the WSFC are the regularisation coefficient for the weight means and the batch size used for training the contour generators. The distribution of the strengths per function for the~\tx{Morlec} database for varying batch size and regularisation coefficient is shown in Fig.~\ref{fig:strengthsperbatch}.

We can see that the batch size impacts significantly the variance of the weights' distributions, as the weight mean regularisation is applied per batch in the backpropagation training. Also, we can see that having no regularisation gives arbitrarily offset strength distributions. 
We perform further experiments with a batch size of 256 and a regularisation coefficient of 10 in order to keep most of the variance while maintaining strong regularisation.

\subsection{Plausibility and performance of WSFC}
\skip -0.4cm
\label{sec:perf}

For assessing the modelling performance, we use the~\tx{Morlec} database, which was chosen for its variety of contour contexts. In this assessment we define the context vector here to be an \emph{attitude indicator}. In fact, it is a one-hot encoding of the attitude of the utterance.
We assess the reconstruction performance of the WSFC and the SFC models
by comparing the root mean square error (RMSE) between the original pitch contour and its reconstruction. 
Pitch is expressed in semitones and only errors within the vocalic nuclei are considered.
We split the data into a training, validation and test sets and trained the WCGs with early stopping. The RMSE results for the test set showed that the WSFC gives a small improvement: 
$1.83\pm1.03$ for the WSFC vs. $1.87\pm1.03$ for
the SFC. This was nevertheless found significant using a paired $t$-test that gave a $p$-value of 
$0.037$.

\subsection{Modelling the impact of attitudes}
\label{sec:att}
Morlec et al~\cite{morlec2001generating} noted that melodic contours of sentences uttered with attitudes such as doubt, evidence, suspicious irony, suggesting a repetition of ``old'' information, exhibited reduced modulations by syntactic functions. 
The WSFC has the capability to contextualise such modulation. \note{For this experiment, we enhanced the context vector to be an \emph{overlap indicator}, which shows not just the attitude, but also any other functional contour appearing within the scope of the current one. This enhancement allows the WSFC to learn a distribution of weights for the functional contours for each attitude, rather than a single value. Note that in this and the following experiments we do not split the data into a train and test set, as we are analysing the expressive capacity of our model in its ability to capture prominence.}

\begin{figure*}[t]
 \centering
 \includegraphics[height=.27\linewidth]{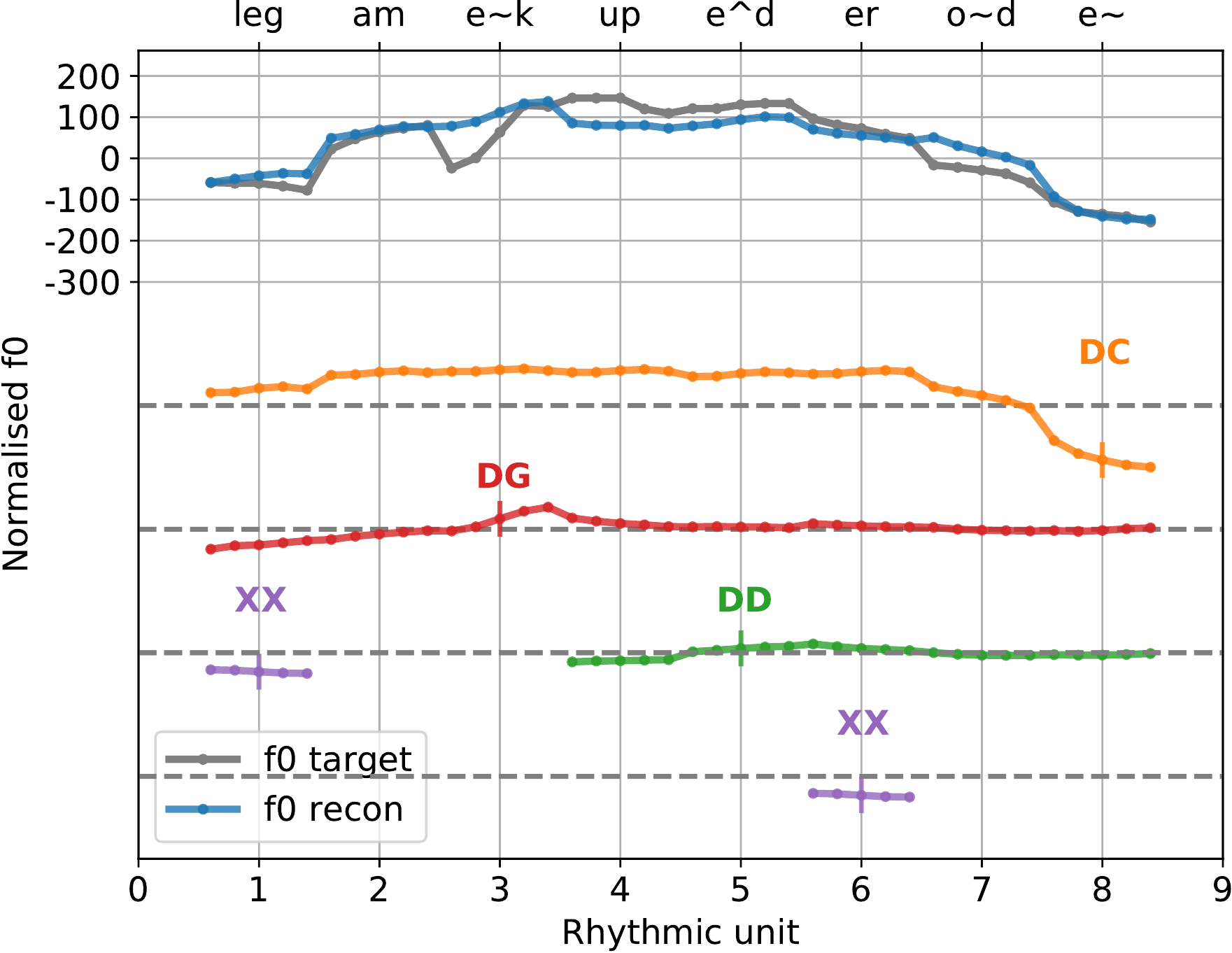}
 \hfill
 \includegraphics[height=.27\linewidth]{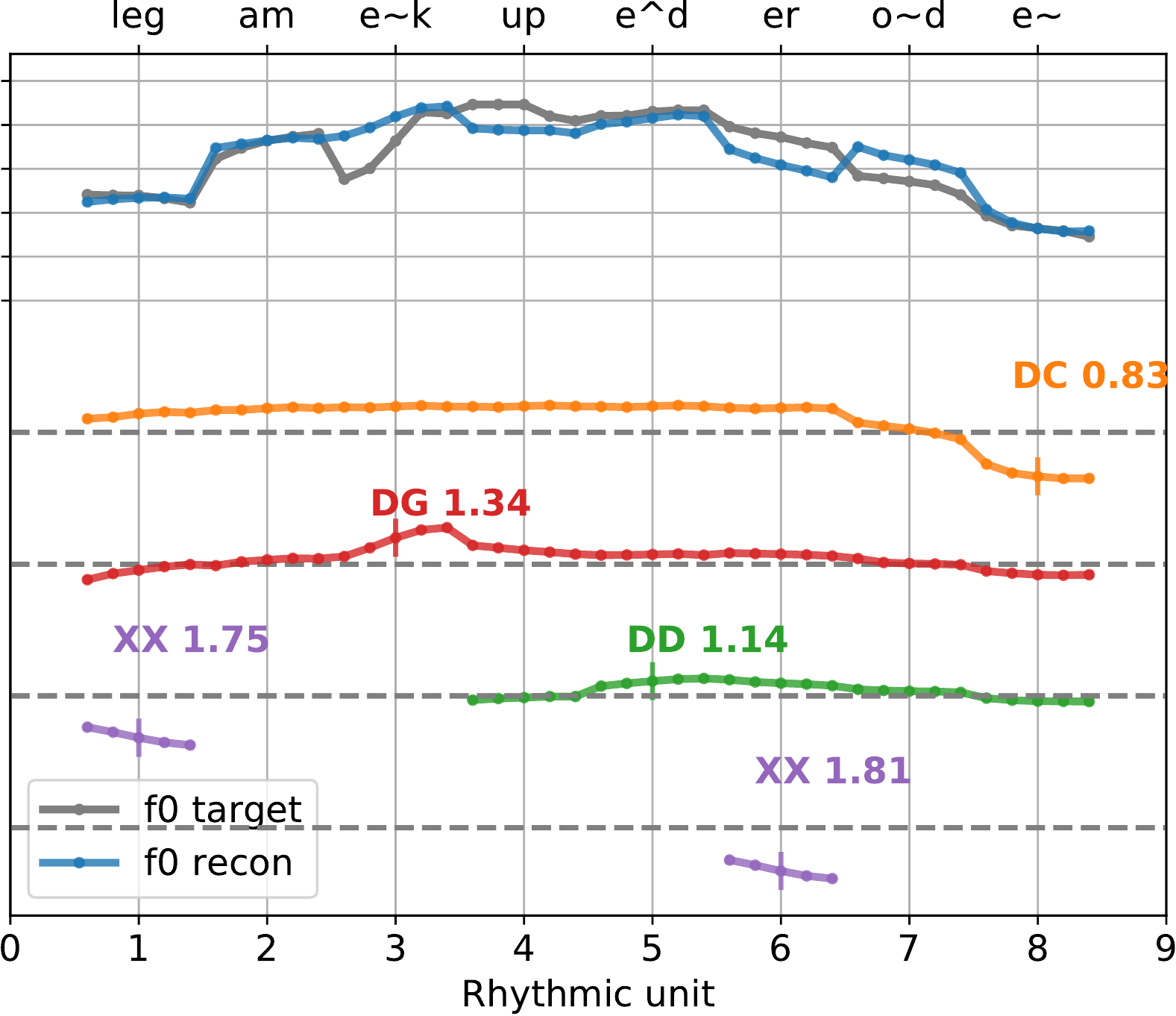}
 \hfill
 \includegraphics[height=.27\linewidth]{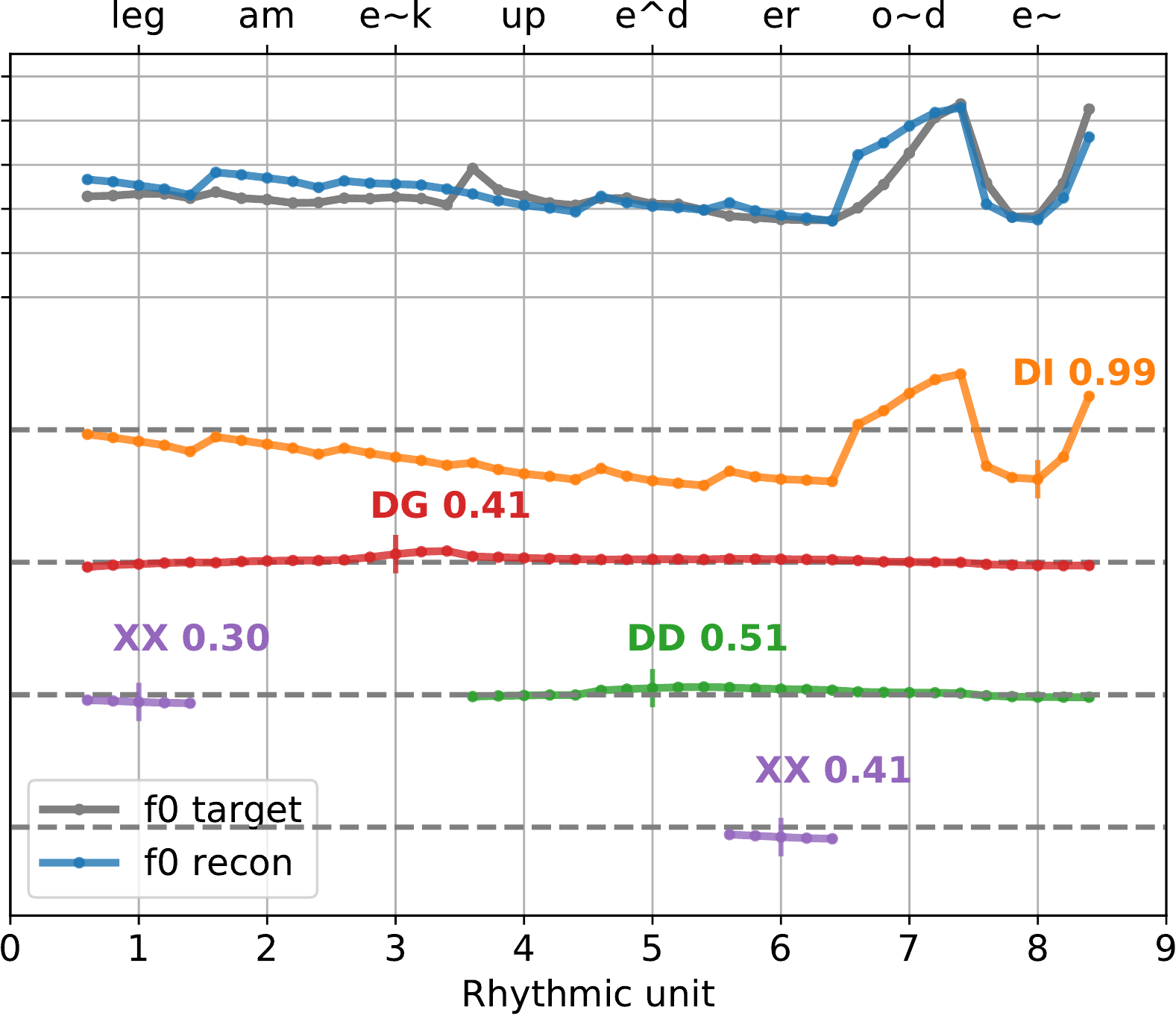}
 \vspace*{-0.2cm}
 \caption{Decomposition of the melody of the French utterance ``Les gamins coupaient des rondins.'' with the SFC (left), and the WSFC for declaration (DC, centre) and incredulous question (DI, right) into constituent functional contours: syntactical dependencies to the left and right (DG, DD), and clitics (XX). Activations of XX, DG and DD are strongly reduced when solicited in the DI context.}
\label{fig:wsfcdecomp}
\end{figure*}

The WSFC decomposition of the same French utterance
for two attitudes: declaration (DC) and incredulous question (DI) is shown in Fig.~\ref{fig:wsfcdecomp}.
We clearly see that the WSFC has captured the quasi-suppression of the contours implementing the DG, DD and XX functions in the context of DI with respect to the full-blown contours in DC. In fact, the weighting factors of XX are reduced by a factor of 5, i.e. from $1.75/1.81$ down to $0.3/0.41$.
This quasi-suppression caused the SFC to average out a smaller clitic contour than the WSFC as is evident in the figure.
The same quasi-suppression phenomenon is also present in all of the other attitudes except exclamation, but it is not shown here for brevity.

\begin{figure}[t]
 \centering
 \includegraphics[height=.32\columnwidth]{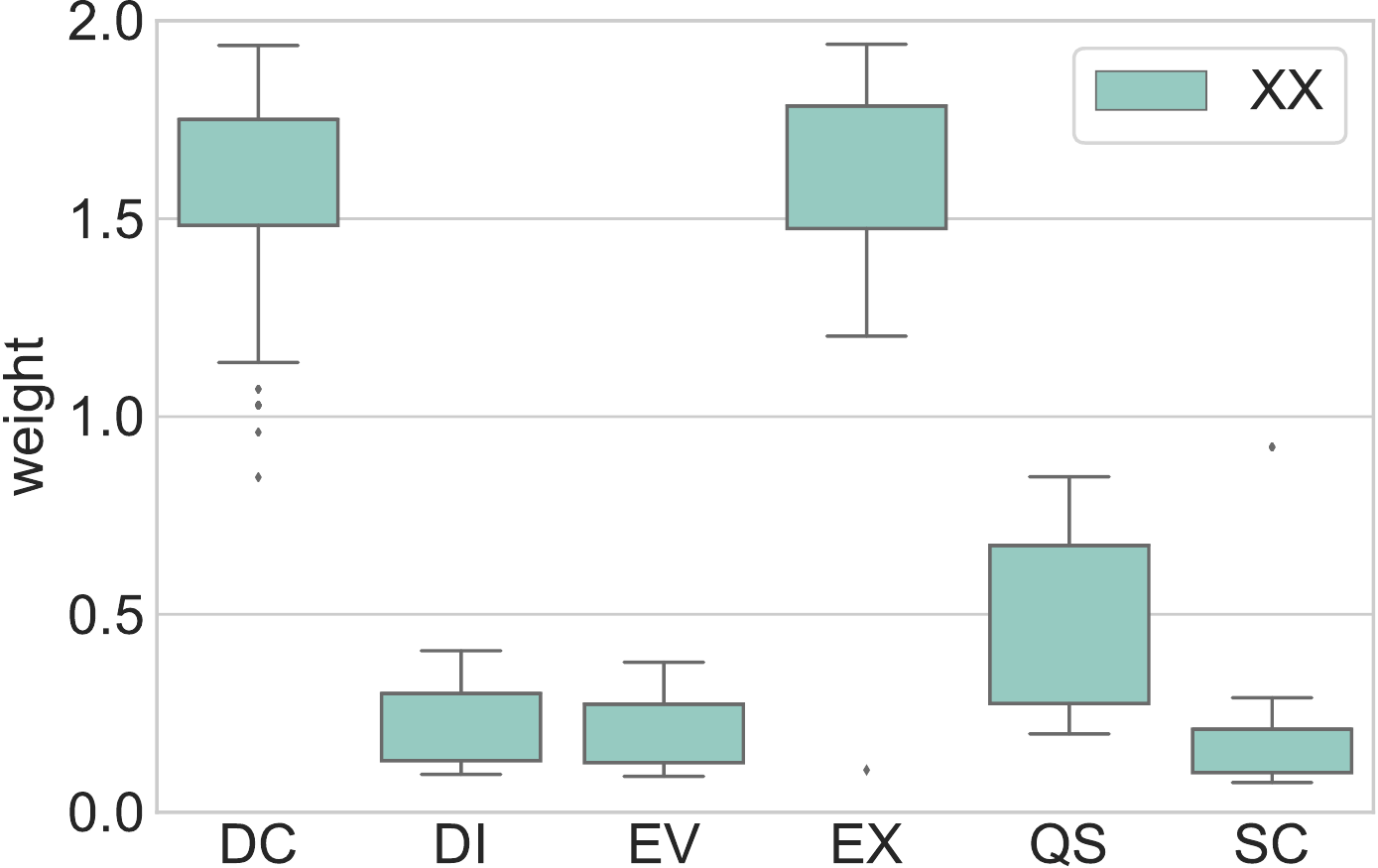}
 \hfill
 \includegraphics[height=.32\columnwidth]{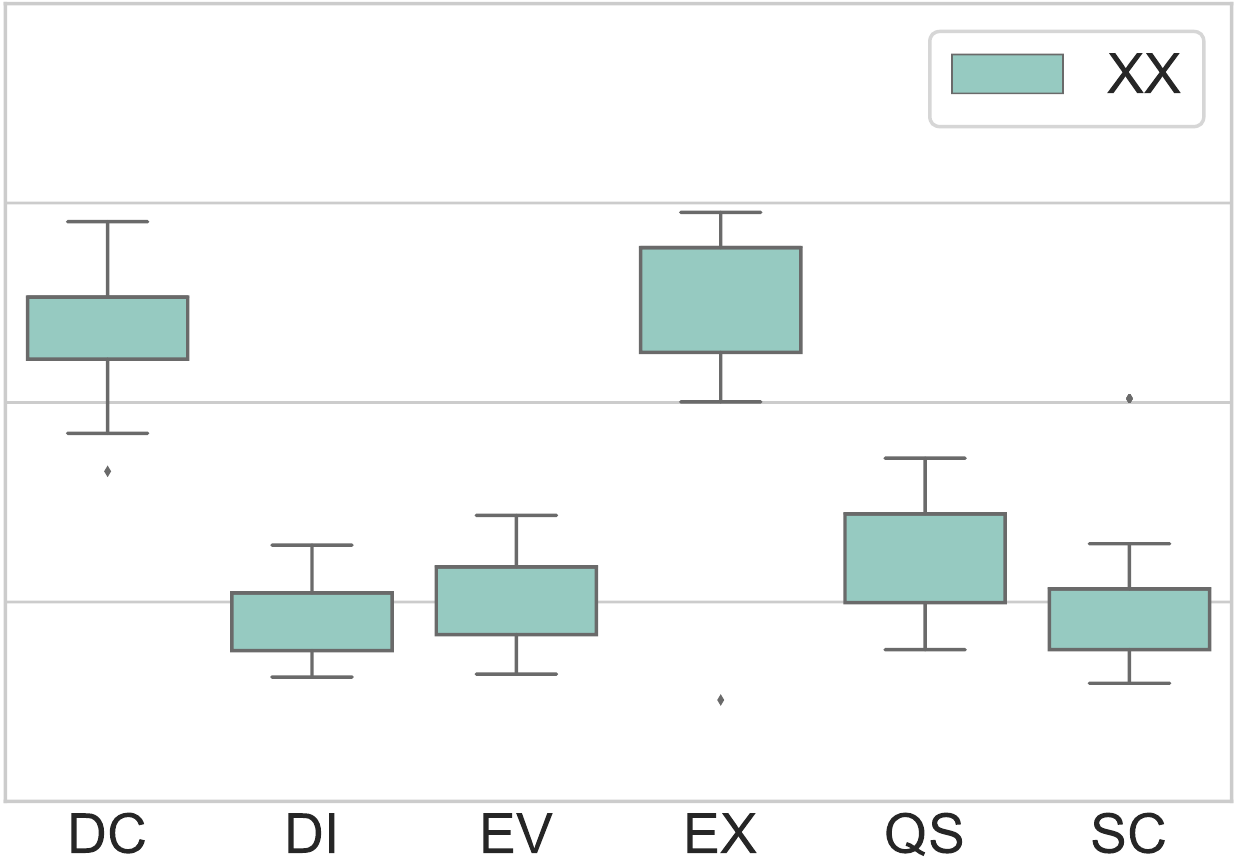}

 \includegraphics[height=.32\columnwidth]{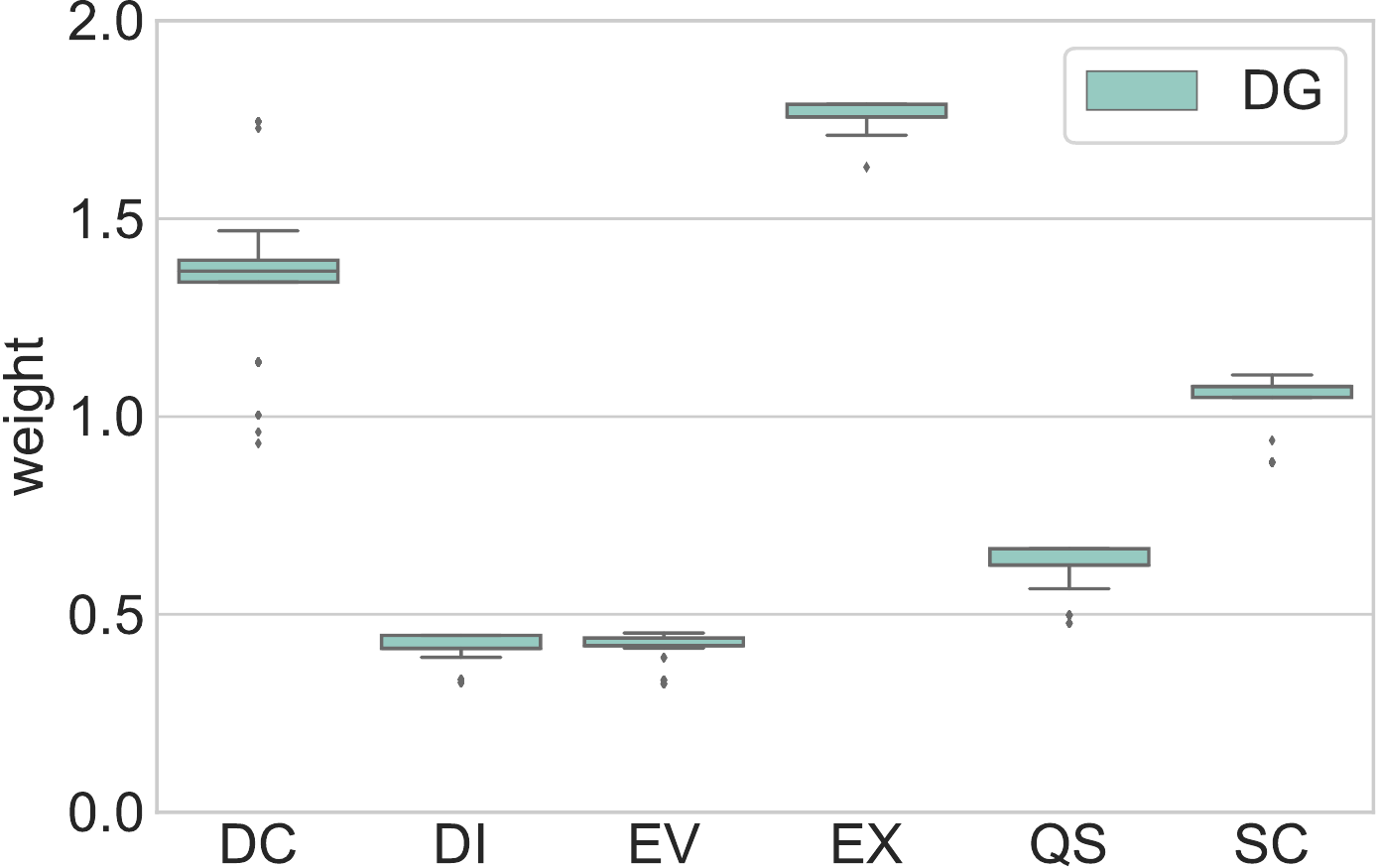}
 \hfill
 \includegraphics[height=.32\columnwidth]{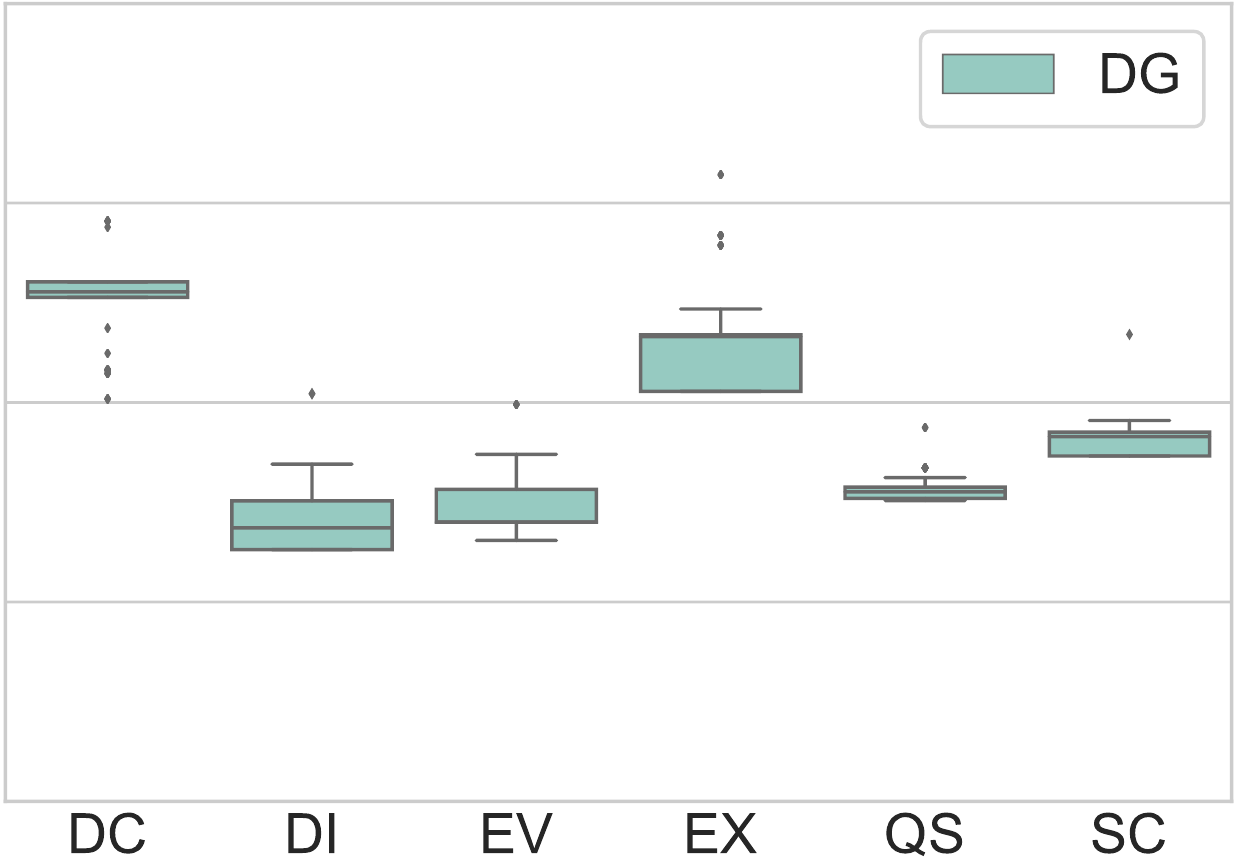}
 \vspace*{-0.2cm}
 \caption{Distribution of WSFC weights for clitic (XX) and right dependency (DG) in \tx{Morlec} without pretrained contours (left column) and with (right column).}
\label{fig:weigthsall}
\end{figure}

We compare here two training strategies:
\emph{full training} -- performed using \tx{Morlec}, and
\emph{pretraining and freezing} -- pretraining the contour generators on DC first, as we suppose it exhibits the full blown syntactic contours. Then freezing the parameters of the contour generator modules when performing the full training. This mirrors the back-off strategy that was followed when using the SFC.
Fig.~\ref{fig:weigthsall} shows the distribution of weights for two functional contours: clitic and left-dependency, with and without pretraining. We can see that
there is a suppression effect by some attitudes on the syntactic contours. We can also see that the WSFC is able to extract a more or less similar distribution of the weights with or without using pre-training. The decomposition proposed by WSFC is thus faithful and robust.
There full training exhibits, however, a stronger contrast. The pretraining causes the final weights distributions for DC and EX to be close to $1$. On the other hand, the distributions obtained with the full training are further spread out, whilst still maintaining the average value of $1$ imposed by the regularisation.

\subsection{Modelling the impact of emphasis}
\label{sec:emph}

The first female speaker subset from \tx{Liu} is used to analyse the impact of emphasis on the Chinese tones. Since \tx{Liu} is designed to minimise tonal modulations by using utterances comprising words with identical tones, we use \tx{Chen} to pretrain the WSFC tone contours with an extended right scope, and then retrain only their strengths on~\tx{Liu}. This is to allow the CGs to capture the well-known carry-over effect in Chinese tones~\cite{xu1997contextual}, thus increasing modelling performance~\cite{gerazov2018tones}. 

For this experiment, we define a further enhanced context vector that is an \emph{emphasis indicator}. It not only includes information on: attitude, word boundary, and emphasis, it also takes into account at which temporal location the function occurs relative to emphasis. This is done because emphasis in Mandarin involves both on-focus expansion of pitch range and post-emphasis compression of pitch range~\cite{liu2005parallel}. Thus the context vector encodes if the RU occurs with no emphasis (None), with emphasis but precedes the final emphasised RU (EMp), is the final emphasised RU (EM), or succeeds it (EMc). 

\begin{figure}[t]
 \centering
 \includegraphics[width=\linewidth]{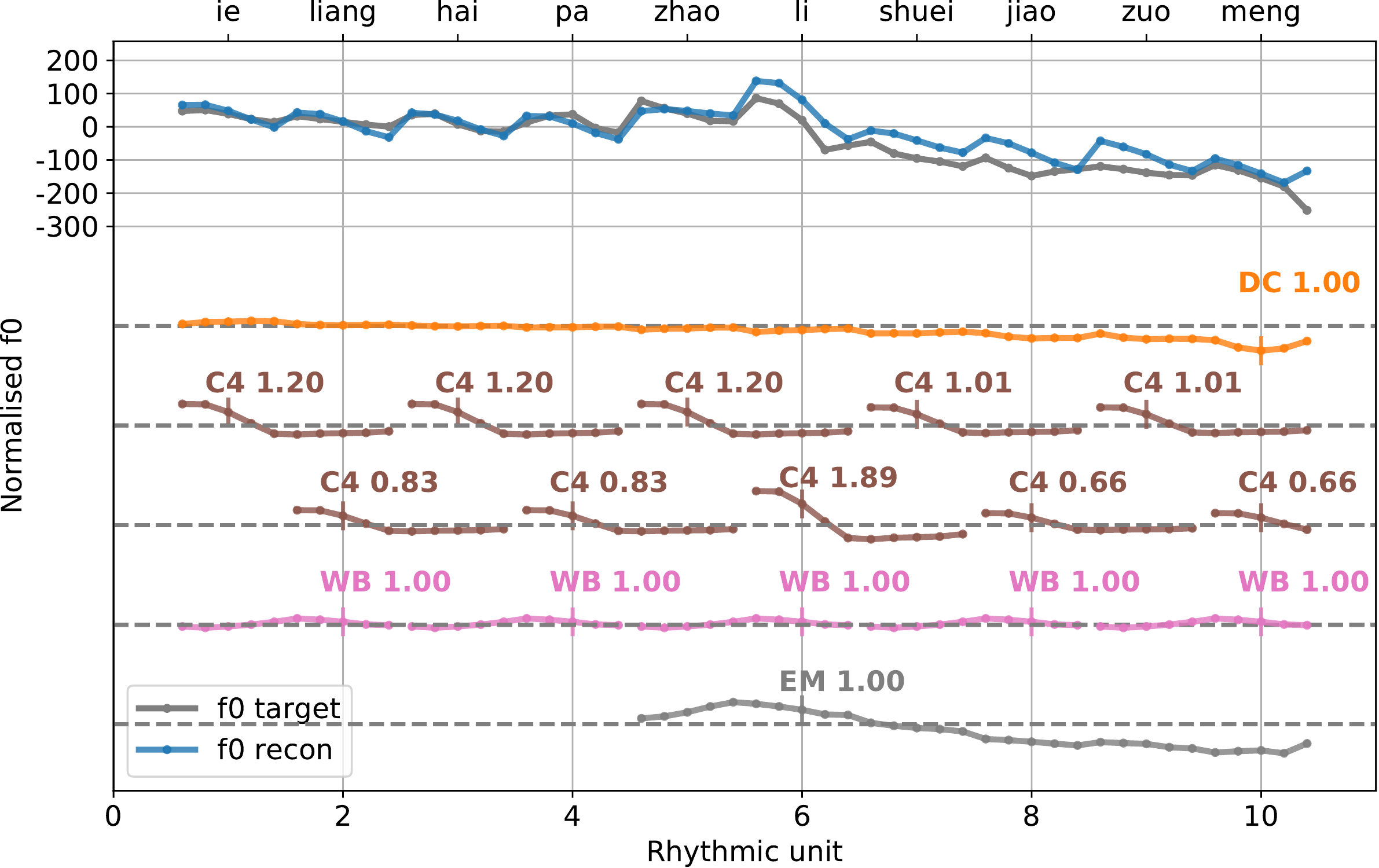}
 \vspace*{-0.6cm}
 \caption{WSFC decomposition of the intonation of the Chinese utterance: ``Ye Liang hai pa \textbf{Zhao Li} shui jiao zuo meng.'', into component contours: declaration (DC), tone contour 4 (C4), word boundary (WB), and emphasis (EM).}
\label{fig:emphasisdecomp}
\end{figure}

An example WSFC decomposition of a Chinese utterance with emphasis is shown in Fig.~\ref{fig:emphasisdecomp}. We can see that the final tone of the second emphasized word ``Li'' is considerably amplified compared to the preceding and following ones. Moreover, we can see that the tones following the emphasis are conversely reduced and that they are also affected by the word boundaries. Finally, we can see that the emphasis contour features an increase in pitch followed by a post-emphasis lowering.

\begin{figure}[!t]
 \centering
 \includegraphics[width=\linewidth]{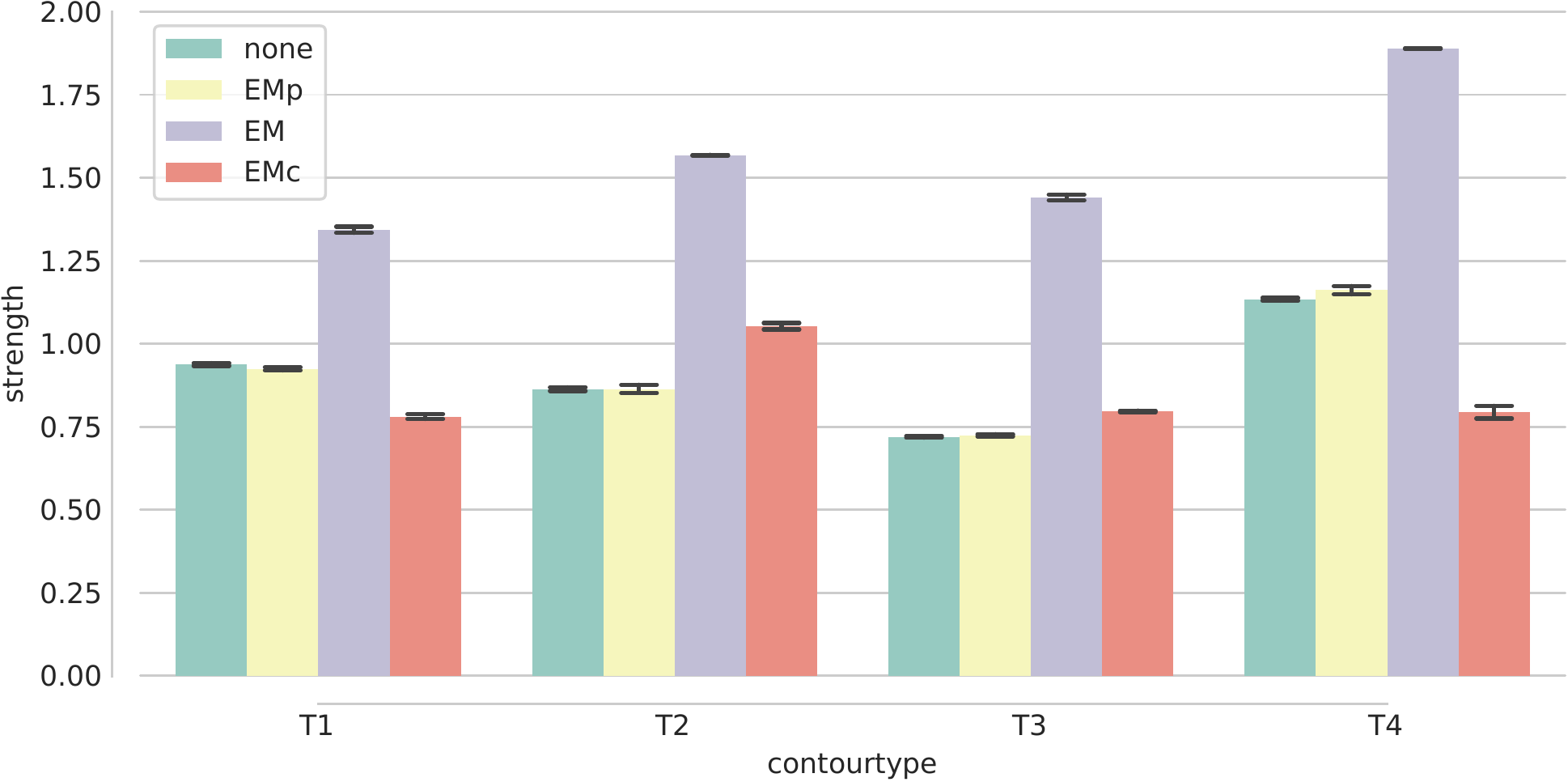}
 \vspace*{-0.6cm}
 \caption{Distribution of WSFC weights in \tx{Liu} 
 with emphasis context.}
\label{fig:emphasis}
\end{figure}

The distribution of weights for the four tones as a function of their placement relative to 
the emphasis are shown in Fig.~\ref{fig:emphasis}. We can see that there is a considerable difference in prominence of the tones located at the final RU in the emphasis. There is also a difference between pre- and post-emphasis weighting for the tones, with weights decreasing for tones 1 and 4 and increasing for tones 2 and 3. It is these changes that, together with the emphasis contour, capture the complex effect of post-emphasis on pitch range dynamics~\cite{liu2005parallel}.

\section{Conclusions}
We proposed a prosody model capable of capturing the prominence of elementary prosodic contours that are based on their context of use. The WSFC has been also shown to improve the modelling performance of the SFC due to an added weighting mechanism. We have demonstrated its robustness and its usefulness in analysing the impact of attitudes and emphasis on prominence in French and Chinese. The described methodology can be used to analyse other effects of context on prominence. Moreover, the proposed WSFC architecture allows for task-specific contextual inputs. We are currently exploring realizations of attitudes in several other languages. 

\section{Acknowledgements}
This work has been conducted with the support of the EU Marie Sk\l{}odowska-Curie Actions Individual Fellowship Project H2020-MSCA-IF-2016 ``ProsoDeep: Deep understanding and modelling of the hierarchical structure of Prosody''.

\pagebreak
\bibliographystyle{IEEEtran}
\bibliography{refs}

\end{document}